\documentclass[preprint]{aastex}
\usepackage{amsmath}
\usepackage{apjfonts}
\usepackage{graphicx}
\usepackage{natbib}

\newcommand{\A}{HESS~J1800$-$240A}
\newcommand{\B}{HESS~J1800$-$240B}
\newcommand{\C}{HESS~J1800$-$240C}
\newcommand{\FGLB}{2FGL~J1800.8$-$2400}
\newcommand{\FGLC}{2FGL~J1758.8$-$2402c}
\newcommand{\W}{Source~W}

\newcommand{\hii}{\ion{H}{2}}

\slugcomment{To be submitted to \apj: v6\_3, March 25, 2014}

\shorttitle{Detailed Investigation of the Gamma-Ray Emission Around W28}
\shortauthors{Hanabata et al.}

\begin{document}


\title{Detailed Investigation of the Gamma-Ray Emission in the\\
 Vicinity of SNR W28 with FERMI-LAT}



\author{
Y.~Hanabata\altaffilmark{2,1}, 
H.~Katagiri\altaffilmark{3,1}, 
J.W.~Hewitt\altaffilmark{4,5}, 
J.~Ballet\altaffilmark{6}, 
Y.~Fukazawa\altaffilmark{7}, 
Y.~Fukui\altaffilmark{8}, 
T.~Hayakawa\altaffilmark{8}, 
M.~Lemoine-Goumard\altaffilmark{9,10}, 
G.~Pedaletti\altaffilmark{11}, 
A.~W.~Strong\altaffilmark{12}, 
D.~F.~Torres\altaffilmark{11,13}, 
R.~Yamazaki\altaffilmark{14}
}
\altaffiltext{1}{Corresponding authors: Y.~Hanabata, hanabata@icrr.u-tokyo.ac.jp; H.~Katagiri, katagiri@mx.ibaraki.ac.jp.}
\altaffiltext{2}{Institute for Cosmic-Ray Research, University of Tokyo, 5-1-5 Kashiwanoha, Kashiwa, Chiba, 277-8582, Japan}
\altaffiltext{3}{College of Science, Ibaraki University, 2-1-1, Bunkyo, Mito 310-8512, Japan}
\altaffiltext{4}{CRESST, University of Maryland, Baltimore County, Baltimore, MD 21250, USA}
\altaffiltext{5}{NASA Goddard Space Flight Center, Greenbelt, MD 20771, USA}
\altaffiltext{6}{Laboratoire AIM, CEA-IRFU/CNRS/Universit\'e Paris Diderot, Service d'Astrophysique, CEA Saclay, 91191 Gif sur Yvette, France}
\altaffiltext{7}{Department of Physical Sciences, Hiroshima University, Higashi-Hiroshima, Hiroshima 739-8526, Japan}
\altaffiltext{8}{Department of Physics and Astrophysics, Nagoya University, Chikusa-ku Nagoya 464-8602, Japan}
\altaffiltext{9}{Centre d'\'Etudes Nucl\'eaires de Bordeaux Gradignan, IN2P3/CNRS, Universit\'e Bordeaux 1, BP120, F-33175 Gradignan Cedex, France}
\altaffiltext{10}{Funded by contract ERC-StG-259391 from the European Community}
\altaffiltext{11}{Institut de Ci\`encies de l'Espai (IEEE-CSIC), Campus UAB, 08193 Barcelona, Spain}
\altaffiltext{12}{Max-Planck Institut f\"ur extraterrestrische Physik, 85748 Garching, Germany}
\altaffiltext{13}{Instituci\'o Catalana de Recerca i Estudis Avan\c{c}ats (ICREA), Barcelona, Spain}
\altaffiltext{14}{Department of Physics and Mathematics, Aoyama Gakuin University, Sagamihara, Kanagawa, 252-5258, Japan}



\begin{abstract}

We present a detailed investigation of the $\gamma$-ray emission in the vicinity of the supernova remnant (SNR) W28 (G6.4$-$0.1) observed by the Large Area Telescope (LAT) onboard the {\it Fermi Gamma-ray Space Telescope}. We detected significant $\gamma$-ray emission spatially coincident with TeV sources \A, B, and C, located outside the radio boundary of the SNR. 
Their spectra in the 2--100~GeV band are consistent with the extrapolation of the power-law spectra of the TeV sources.
We also identified a new source of GeV emission, dubbed \W, which lies outside the boundary of TeV sources and coincides with radio emission from the western part of W28.
All of the GeV $\gamma$-ray sources overlap with molecular clouds in the velocity range from 0 to 20~km~s$^{-1}$.
Under the assumption that the $\gamma$-ray emission towards \A, B, and C comes from $\pi^0$ decay due to the interaction between the molecular clouds and cosmic rays (CRs) escaping from W28, they can be naturally explained by a single model in which the CR diffusion coefficient is smaller than the theoretical expectation in the interstellar space.
The total energy of the CRs escaping from W28 is constrained through the same modeling to be larger than $\sim$~2~$\times$~10$^{49}$~erg.
The emission from \W~can also be explained with the same CR escape scenario.
\end{abstract}


\keywords{acceleration of particles --- cosmic rays --- diffusion --- ISM: supernova remnant --- gamma-rays: ISM}



\section{Introduction}

Diffusive shock acceleration (DSA) operating at the shocks of supernova remnants~\citep[SNRs;][and references therein]{Reynolds08} is the most likely mechanism to convert the kinetic energy released by supernova explosions into high energy particles (cosmic rays; CRs) that obey a power-law type distribution.
Evidence of the CR proton acceleration in SNRs has emerged from the detection of GeV $\gamma$ rays from some SNRs interacting with molecular clouds such as W51C, W44, and IC 443~\citep{W51C,W44,IC443} by the Large Area Telescope (LAT) on board the {\it Fermi  Gamma-ray Space Telescope}.
The intense GeV emission from those SNRs is naturally explained by $\pi^0$ decay produced in inelastic collisions of the accelerated protons with dense gas.
This was recently confirmed by the detection of the characteristic spectral feature produced by the decay of $\pi^0$s in W44 and IC 443~\citep{W44_AGILE, W44_Science}.
In DSA theory, CRs accelerated at the shock are scattered by self-generated magnetic turbulence. 
Since the highest-energy CRs in the shock precursor at the upstream are prone to lack self-generated turbulence, they are expected to escape from the shock~\citep{Ptuskin03}.
However, it has been unclear how the CRs escape from SNRs and propagate into the interstellar medium (ISM) because the interplay among the CRs, the  magnetic turbulence, and the surrounding environment of SNRs is not well understood.

If an SNR is in a dense environment, we can expect an enhancement of the $\pi^0$-decay $\gamma$ rays from molecular clouds illuminated by the escaping CRs in the vicinity of the SNR~\citep{Aharonian96,Gabici07}. 
For example, the $\gamma$-ray emissions near middle-aged SNRs G8.7$-$0.1 and W44 are naturally explained by the above scenario~\citep{G8.7,Uchiyama12}.
The energy dependence of the diffusion coefficient of the CRs alters their spectrum, which affects the spectral shape of the resulting $\gamma$-ray emission~\citep{Aharonian96,Gabici09,Ohira11}.
Thus, we can constrain the diffusion coefficient by measuring the wide-band $\gamma$-ray spectrum of the emission around SNRs.

W28 (also known as G6.4$-$0.1) is a mixed-morphology SNR whose age is estimated to be (3.5--15)~$\times$~10$^4$~yr~\citep{Kaspi93}. In this paper, we adopted the same age of 4~$\times$~10$^4$~yr as used in~\citet{W28}. 
The SNR is located within a molecular cloud complex with a mass of 1.4~$\times$~10$^{6}$~$M_\odot$~\citep{Reach05} and interacts with some parts of the cloud, traced by the detection of OH (1720~MHz) masers~\citep{Frail94,Claussen97,Claussen99}.
Observations of molecular lines placed W28 at a distance of $\sim$1.9~kpc~\citep{Velazquez02}.
GeV $\gamma$-ray emission associated with W28 has been detected by the LAT and the Gamma-Ray Image Detector (GRID) onboard {\it AGILE}~\citep{Tavani08}. A natural explanation is the decay of $\pi^0$s due to the interaction of the cloud and CRs accelerated in the SNR~\citep{W28,AGILE_W28}. 
W28 is considered to have entered the radiative phase~\citep{Lozinskaya92} as indicated by optical filaments~\citep{Lozinskaya74}. 
Thus, we can expect that CRs have escaped into the surrounding ambient medium.

H.E.S.S. observations of W28 have revealed four TeV $\gamma$-ray sources~\citep{Aharonian08}: HESS~J1801$-$233, located along the northeastern boundary of W28, and a complex of three sources, \A, B, and C, located to the south, outside the radio boundary. 
The southern H.E.S.S. sources spatially correspond with molecular clouds whose distances are consistent with that of W28~\citep{Aharonian08}, suggesting the possibility that their origins are due to runaway CRs from the SNR.
Thus, this region is one of the best sites to study CR diffusion.
Although \cite{W28} detected only one source associated with \B\ in the first year of observations, there are two LAT sources in the southern region listed in the second {\it Fermi}-LAT catalog~\citep[2FGL;][]{2FGL}. 
If the GeV and TeV emissions originate from CRs escaping from W28, we can constrain the diffusion coefficient of the particles in this region.

In this paper, we report a detailed analysis of the LAT sources surrounding W28, based on 4 years of data.
First, we give a brief description of the observations and data selection in Section~\ref{obs}. The analysis procedure and results are presented in Section~\ref{ana_res}, along with the spectra of the LAT sources. The discussion is given in Section~\ref{discussion} followed by conclusions in Section~\ref{conclusion}.

\section{Observations and Data Reduction}
\label{obs}

The LAT is the main instrument of {\it Fermi}, detecting $\gamma$ rays by conversion into electron-positron pairs in the energy range from $\sim$~20~MeV to $>$~300~GeV~\citep{Atwood09}. It contains a high-resolution converter/tracker for direction measurement of the incident $\gamma$ rays, a CsI (TI) crystal calorimeter for energy measurement, and an anti-coincidence detector to identify the background of charged particles. 
The LAT has a large effective area ($\sim$~7500~cm$^2$ on-axis for $>$~2~GeV), a wide field of view ($\sim$~2.4~sr), and a good point-spread function (PSF; the 68\% containment angle at $>$~2~GeV is smaller than $\sim$~0$\fdg$6).
The on-orbit calibration, event classification and instrument performance are described in detail in~\citet{Abdo09}.

We have analyzed 4 years of data within the energy range 2--100~GeV in the vicinity of W28, collected from 2008 August 4 to  2012 August 18, with a total exposure of $\sim$~1.2~$\times$~10$^{11}$~$\rm cm^2~s$ at 2~GeV. The LAT was operated in sky-survey mode nearly continuously. In this observing mode, the LAT scans the whole sky every two orbits ($\sim$~3~hr), obtaining complete sky coverage and approximately uniform exposure.

We used the standard LAT analysis software, {\it ScienceTools} version v9r30, publicly available from the {\it Fermi} Science Support Center (FSSC)\footnote{Software and documentation of the {\it Fermi ScienceTools} are distributed by {\it Fermi} Science Support Center at http://fermi.gsfc.nasa.gov/ssc.}. We used the post-launch instrument response functions (IRFs) P7V6~\citep{Ackermann12}  and applied the following event selection criteria: (1) events should be classified as $``$Source$"$  class, (2) the reconstructed zenith angles of the arrival direction of $\gamma$ rays should be smaller than 100$^\circ$ to minimize the contamination from Earth-limb $\gamma$-ray emission, and (3) only time intervals when the center of the LAT field of view is within 52$^\circ$ of the local zenith, are accepted to further reduce the contamination by Earth's atmospheric emission. 
The $\gamma$-ray burst GRB~100826A~\cite[]{McEnery10} occurred within the region used for the analysis in this paper. However the event data are not included because they do not satisfy the above criteria (3). 
Thus, we did not need to apply any additional time cut.

Two different tools were used to perform the spatial and spectral analysis: {\tt gtlike} (in binned mode) and {\tt pointlike}. {\tt gtlike} is a standard maximum-likelihood method~\citep{Mattox96}. {\tt pointlike} is an alternate binned likelihood technique, optimized for characterizing the extension of a source, that has been extensively tested against {\tt gtlike}~\citep{Lande12}.

\section{Analysis and Results}
\label{ana_res}
\subsection{Morphological Analysis and Source Position}
\label{image_ana}

\begin{figure}
\epsscale{.8}
\plotone{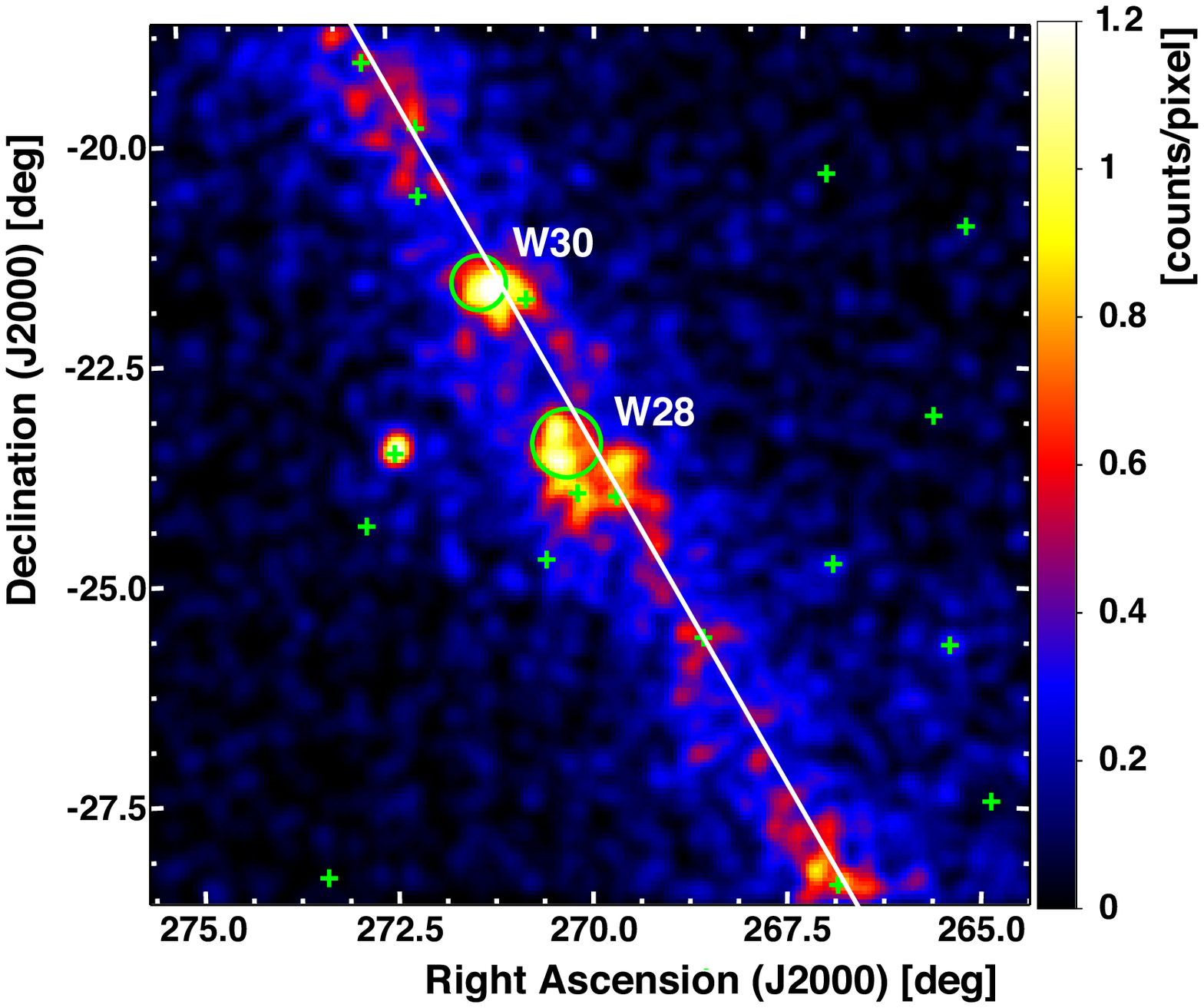}
\caption{\textit{Fermi} LAT 10--100~GeV count map around W28. The count map is smoothed by a Gaussian kernel of $\sigma~=~0\fdg20$. The pixel size is $0\fdg05$. 
The green circles and crosses indicate the extended and point sources in the 2FGL catalog~\citep{2FGL}, respectively.
The white line from top left to bottom right indicates the Galactic plane.}
\label{lat_cmap_wide}
\end{figure}

Here, we analyzed the morphology of high energy $\gamma$-ray emission in the vicinity of W28.
We made a counts map in the 10--100~GeV energy band to take advantage of optimal angular resolution and weaker Galactic diffuse emission. Figure~\ref{lat_cmap_wide} shows the map in a 10$^\circ \times 10^\circ$ region around W28.
Emission around W28 can be clearly seen against the Galactic diffuse emission.
There are two LAT sources in the vicinity of W28 in the 2FGL catalog~\cite[]{2FGL}: \FGLC~and \FGLB.

\begin{figure}
\epsscale{1.0}
\plotone{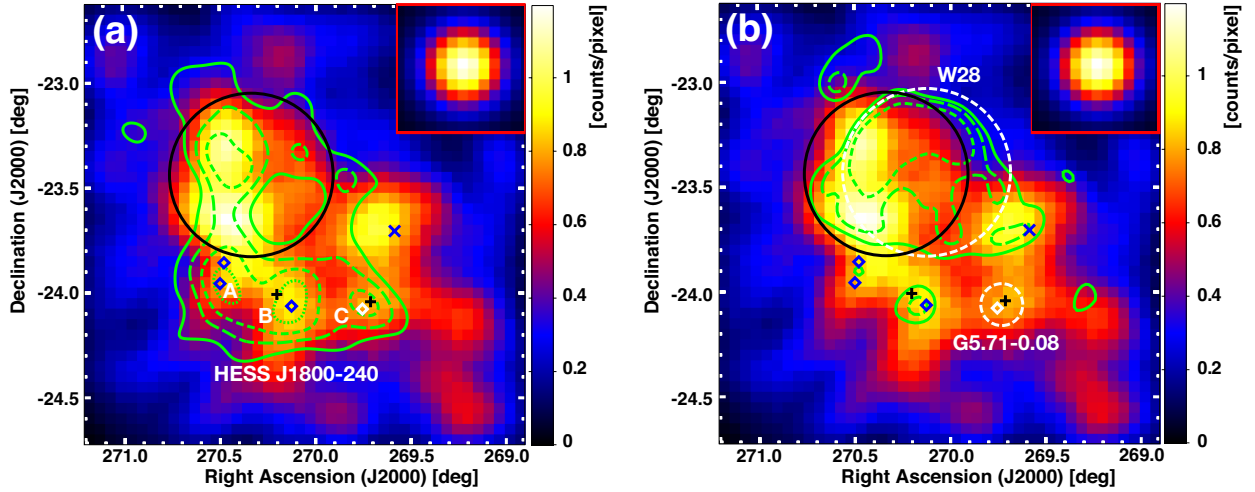}
\caption{
The LAT count map between 10 and 100~GeV around W28 superimposed on (a) H.E.S.S. and (b) VLA contours.
The maps have a pixel size of 0$\fdg$05 and are smoothed by a Gaussian kernel of $\sigma = 0\fdg20$. 
The inset of each figure shows the simulated LAT PSF with a photon index of 2.5 in the same energy range, adopting the same smoothing. 
Locations of 2FGL catalog sources included in the emission model are indicated with black marks; circle for W28, left-hand cross for \FGLB, and right-hand one for \FGLC, respectively. A blue x indicates the best-fit position of \W. Blue diamonds on the left indicate \hii~regions: G6.1$-$0.6~\citep{Kuchar97} and G6.225$-$0.569~\citep{Lockman89}. The blue diamond on the right is W28A2 (see the text). The white diamond indicates the OH maser spot associated with G5.71$-$0.08~\citep{Hewitt09}.
Green contours in panel (a) show the H.E.S.S. significance map for TeV $\gamma$ rays at 20, 40, 60, and 80\% levels~\citep{Aharonian08}. Bright TeV spots in the south are \A, B, and C as indicated in the figure. Green contours in panel (b) indicate the VLA~90~cm image at 5, 10, and 20\% of the peak intensity~\citep{Brogan06}. Outer boundaries of SNR W28 and G5.71$-$0.08, as determined from the radio images, are drawn as white dashed circles.}
\label{lat_cmap}
\clearpage
\end{figure}

Figure~\ref{lat_cmap} shows a close-up view of the LAT counts map, superimposed on the H.E.S.S. significance map~\citep{Aharonian08} and the radio image by the Very Large Array~\citep[VLA;][]{Brogan06}. 
Multiple spatial associations are evident, allowing GeV and TeV emission to be correlated: between the northern part of W28 and HESS J1801$-$233, between  \FGLC\ and \C, and between \FGLB\ and \B~\citep{Aharonian08}.
GeV $\gamma$ rays also overlap with \A\ although there is no 2FGL source there.
We also found an additional GeV source to the west beyond the observed TeV emission. 
Note that this $\gamma$-ray source cannot be seen clearly below 10~GeV.
Here we will refer to the GeV emissions coincident with the two TeV $\gamma$ ray sources as \B\ and 240C, and to the western GeV emission as \W.
In the VLA image, \W~overlaps with the western shell of W28 whereas 240B and 240C overlap with radio emission unrelated to W28: \ion{H}{2} region W28A2 (G5.89$-$0.39) and SNR G5.71$-$0.08~\citep{Brogan06}, respectively.

In order to evaluate the morphology of the GeV emission around W28, we fit models to the data using the maximum likelihood framework with {\tt pointlike}.
The likelihood is the product of the probabilities of observing the $\gamma$-ray counts within each spatial and energy bin for a specified emission model. The best parameter values are estimated by maximizing the likelihood of the data over the set of models~\citep{Mattox96}.
The probability density function for the likelihood analysis includes (1) individual sources detected in the 2FGL catalog within 15$^\circ$ of W28, (2) the Galactic diffuse emission resulting from CR interactions with interstellar medium and radiation based on the LAT standard diffuse background model, {\tt gal\_2yearp7v6.fits} available from the FSSC\footnote{The model can be downloaded from\\ http://fermi.gsfc.nasa.gov/ssc/data/access/lat/BackgroundModels.html}, and (3) an isotropic component to represent extragalactic $\gamma$ rays and residual CR background using a tabulated spectrum written in {\tt iso\_p7v6source.txt} also available from the FSSC. The region of interest for the binned maximum likelihood analysis based on Poisson statistics\footnote{As implemented in the publicly available {\it Fermi} Science Tools. The documentation concerning the analysis tools and the likelihood fitting procedure is available from\\ http://fermi.gsfc.nasa.gov/ssc/data/analysis/documentation/Cicerone/.} was a square region of 16$^\circ \times 16^\circ$ in Galactic coordinates centered on W28 with a pixel size of 0$\fdg$1.

First, we determined the strength of the diffuse $\gamma$-ray emission around W28. To take advantage of the narrower PSF at higher energies, we analyzed data from 2 to 100~GeV. 
To account for any effects from nearby sources on the fit, we set free the normalization and spectral index of a power-law model applied to the Galactic diffuse emission and the locations and spectral normalizations of all 2FGL sources within $4^\circ$ of the direction of W28.
To represent the emission near \W~more accurately, the spectral parameters of W28, \B~and 240C were also set free. 
All parameters of the sources beyond 4$^\circ$ were fixed to the values in the 2FGL catalog. 
After the fit, the spectral parameters except for 240B and 240C were fixed to the values obtained in the above analysis. 
We used a uniform disk model as the spatial template of W28, as in the 2FGL catalog. Fitting the extension and position of W28 gave consistent results with those in the catalog. 
We substituted the H.E.S.S. significance map of HESS~J1801$-$233~\citep{Aharonian08} for the template and found that it cannot represent the GeV emission from W28.

We performed a series of maximum likelihood fits to investigate the GeV morphology around W28, adopting a power-law spectral form for all sources of interest. 
First, we added \W~as a point source and varied its position. As a result, we obtained the resulting maximum likelihood value for three point sources composed by 240B, 240C, and \W~($L_{\rm 3p}$) with respect to that for the no-source hypothesis ($L_0$). The likelihood ratio, $-2\ln$($L_0/L_{\rm 3p}$) (12 degrees of freedom) of $\approx$~335, was substantially better than that obtained for a model containing only the two point sources 240B and C, $-2\ln$($L_0/L_{\rm 2p}$) (8 degrees of freedom) $\approx$~293, where $L_{\rm 2p}$ is the likelihood for the two sources with the optimized positions (see Table~\ref{morphology_result}). We therefore concluded that Source~W is significantly detected. Note that the locations of the other two sources do not significantly differ from those in the 2FGL catalog. 
Then we added a point source at the peak of the TeV emission 240A. The obtained likelihood ratio increased by $\sim$~29, so we concluded that 240A is also significantly detected. 
In addition we substituted the four point sources model with a morphological model composed by \W~in addition to the H.E.S.S. significance map.
The map was divided into individual sources 240A, 240B, and 240C with separate spectral parameters whose boundaries were determined from the apparent TeV morphology.
For the latter we extracted the regions above 4$\sigma$ to avoid the background fluctuations.
The resulting maximum likelihood ratio of $\approx$~383 (11 degrees of freedom) is larger than that of the four point sources model. Thus, we concluded that the best-fit model for the GeV emission is provided by including the H.E.S.S. template and a separate source \W, and used this model for the spectral analysis. 
Note that this result holds even if the extraction threshold for the template extraction from the H.E.S.S. significance map is changed by $\pm$1$\sigma$.

\W~is consistent with a newly detected point source. The upper limit on its radius, assuming a uniform disk spatial model, was 4$'$ at 68\% confidence level. The position of \W\ in J2000 was obtained as (R.A., decl.)~=~(17${}^{\rm h}$58${}^{\rm m}$.2, $-$23${}^\circ$42$'$.3) with an error radius of 0$\fdg$033 at 68\% confidence level. 
We found no other obvious multiwavelength counterparts to GeV sources, such as pulsars and blazars, within the positional error radius of 0$\fdg$054 at the 95\% confidence region.
We tested the possibility that \W~is a background active galactic nucleus, such as a blazar, which typically has a longer variability time scale of the $\gamma$-ray flux than a few months. Although \W~is not bright enough to investigate the variability on monthly time scales, we can expect that the test statistics (TS) of Source W slowly increases each year if the source is steady. Indeed, we found that its TS gradually increases with time.
However, it remains possible that \W~may be a $\gamma$-ray blazar with repetitive yearly activity. We also could not exclude the possibility that \W~is a $\gamma$-ray pulsar due to the lack of photon statistics for a pulsation search.

\subsection{Energy Spectrum}
\label{energy_spectrum}

\begin{figure}
\epsscale{.49}
\plotone{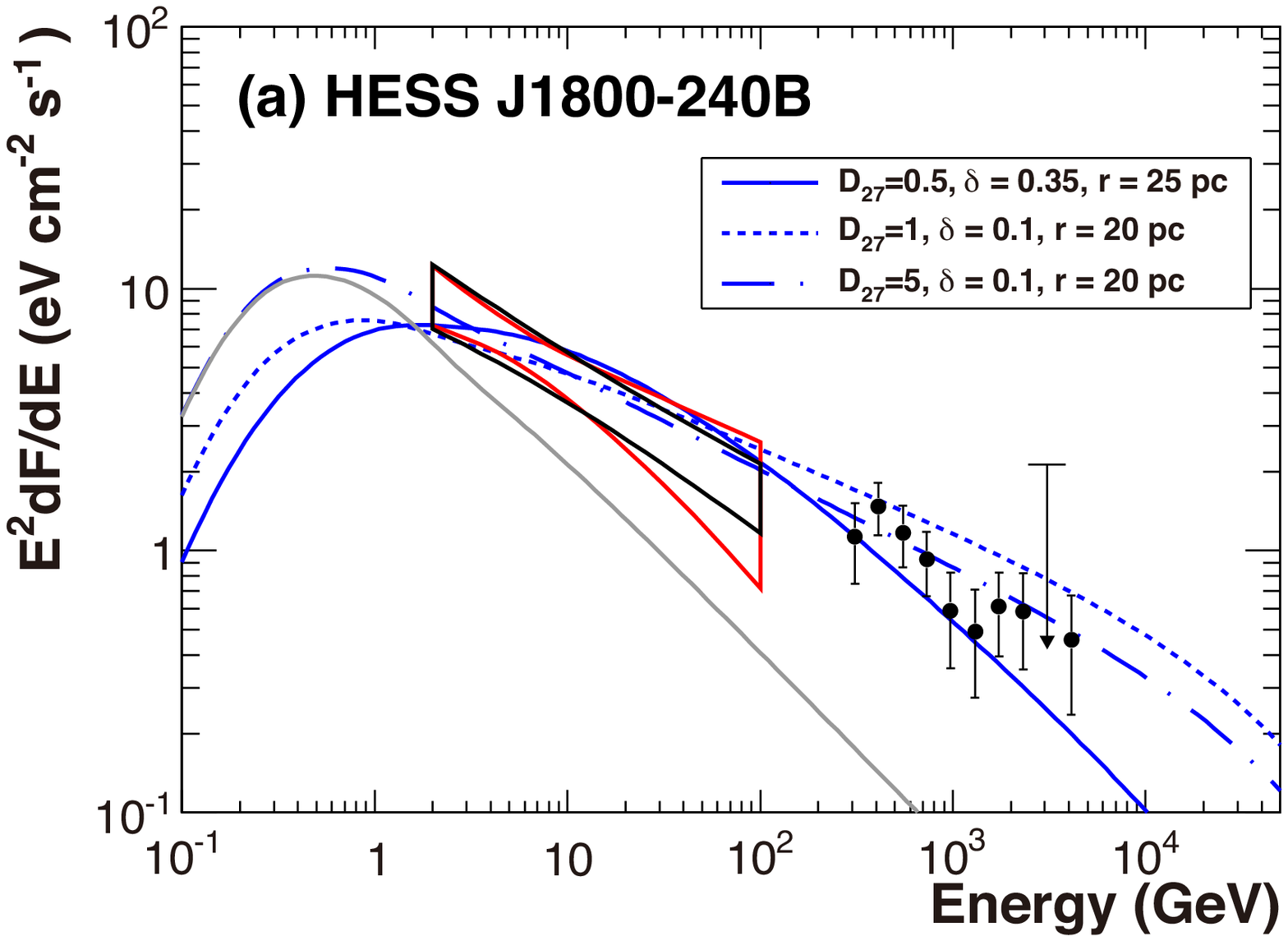}
\plotone{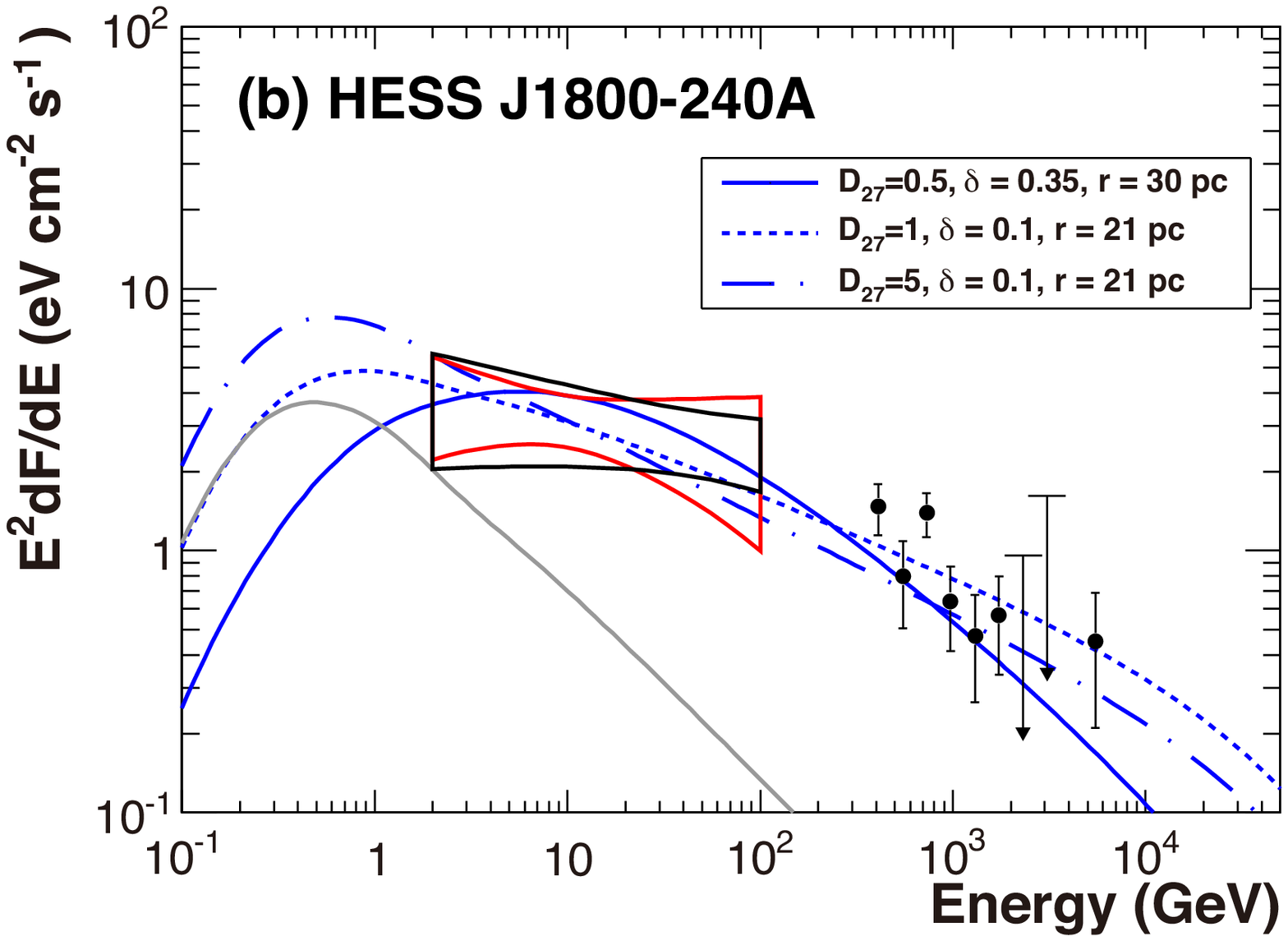}
\plotone{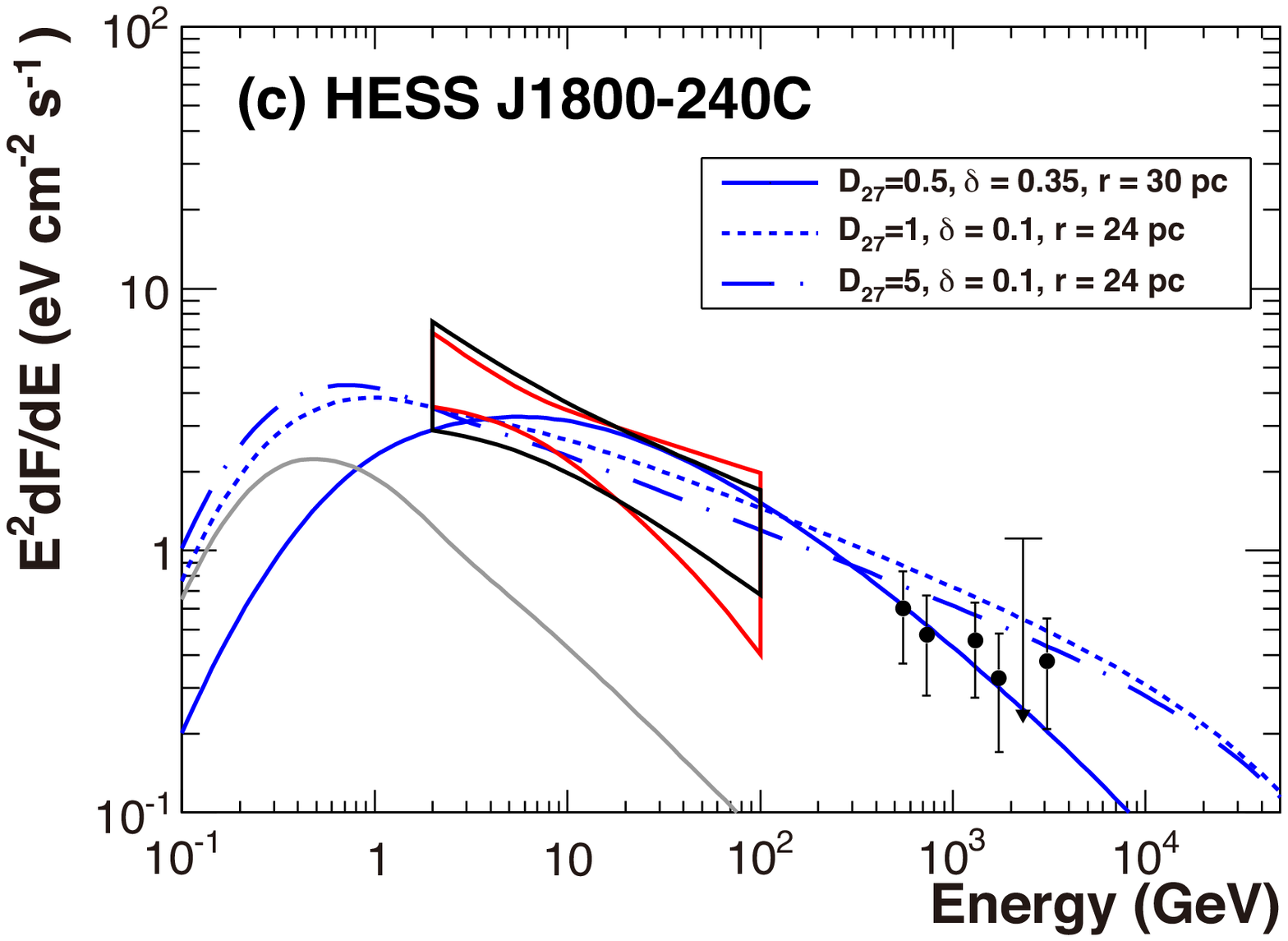}
\plotone{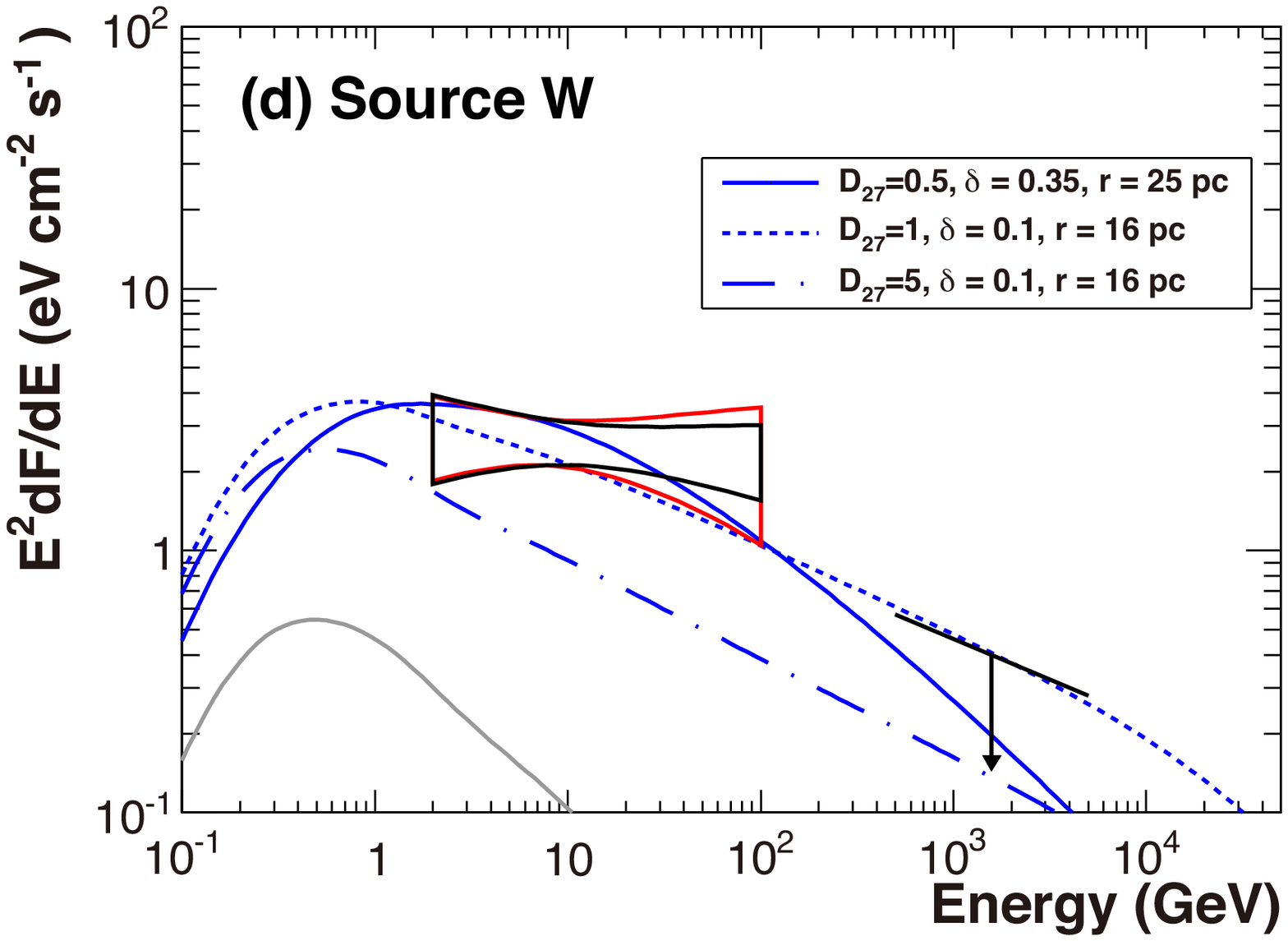}
\caption{ 
The GeV-TeV spectral energy distributions of (a) \B\, (b) 240A, (c) 240C, and (d) \W, respectively. 
The red regions are the 68\% confidence range of the LAT spectra. The black regions show the combined systematic errors. The black circles show the data points for the H.E.S.S. measurement~\citep{Aharonian08}. In panel (b), the best-fit model of 240B obtained by~\citet{Aharonian08} is shown as an upper limit in the TeV range.
In each panel, the blue solid, dotted, and dashed-dotted lines show the model curves with (a) ($D_{27}$, $\delta$, $r$)~=~(0.5, 0.35, 25~pc), (1, 0.1, 20~pc), and (5, 0.1, 20~pc), respectively, (b) ($D_{27}$, $\delta$, $r$)~=~(0.5, 0.35, 30~pc), (1, 0.1, 21~pc), and (5, 0.1, 21~pc), respectively, (c) ($D_{27}$, $\delta$, $r$)~=~(0.5, 0.35, 30~pc), (1, 0.1, 24~pc), and (5, 0.1, 24~pc),respectively, (d) ($D_{27}$, $\delta$, $r$)~=~(0.5, 0.35, 25~pc), (1, 0.1, 16~pc), and (5, 0.1, 16~pc), respectively. 
The grey curve in each panel indicates the upper limit of the $\gamma$-ray emission produced by the sea of Galactic CRs in the same CR-illuminated cloud.}
\label{spec}
\end{figure}

For the spectral analysis of LAT sources in this region, we used the maximum likelihood fit tool, {\tt gtlike}.
Each source was modeled as a simple power-law function $\frac{dN(E)}{dE}~\propto~E^{-\alpha}$ for the fit.
The obtained spectral index of 2.77~$\pm$~0.06 and flux level for W28 were in agreement with the previous result reported by~\cite{W28}, but the spectral energy distribution is not shown here because we focus on the sources near W28.
Figure~\ref{spec} shows the resulting spectral energy distribution for each source along with 68\% confidence regions. We show the best-fit model of \C~obtained by~\cite{Aharonian08} as an upper limit for \W~in the TeV range (see Figure~\ref{spec}(d)).
The obtained spectral indices and TS values are shown in Table~\ref{spec_result}. 
The spectral index of \B~obtained here is consistent within the uncertainties with Source~S in~\cite{W28}.

The LAT spectra of 240B and 240C smoothly connect to the H.E.S.S. measurements, while 240A has a slightly harder spectral index than the value of 2.55~$\pm$~0.18 found by H.E.S.S.~\citep{Aharonian08}.
 Thus, 240A is expected to have a spectral break in the GeV to TeV range.
However, a simple power-law spectrum with a lower flux and softer spectral index at the 1$\sigma$ level away from the LAT best-fit values is consistent with the H.E.S.S. spectrum.
This result is different from that presented in~\cite{W28} where it was shown that 240A has a spectral break in the LAT band.
This might be caused by the difference in source extension or an improved understanding of the diffuse background models.
\W~also can be expected to have a spectral break in the GeV to TeV band from~Figure \ref{spec}(d). We fitted the LAT spectrum with a power-law with an exponential cutoff. However, no clear evidence for a break was found from the GeV data alone.

We considered the systematic errors due to the extraction threshold of the H.E.S.S. significance map, the uncertainty of the GeV $\gamma$-ray morphology of W28, the LAT effective area, and the modeling of interstellar emission.
We evaluated the systematic errors associated with the H.E.S.S. map by changing the nominal threshold of 4~$\sigma$ that we used to extract the morphology templates, as explained in Section~\ref{image_ana}, by $\pm$~1~$\sigma$.
To account for imperfections in the spatial model describing the morphology of W28 (see Figure~\ref{lat_cmap}), we divided the uniform disk into 4 quadrants to represent the morphology more accurately and estimated systematic errors. The uncertainties in the LAT effective area are 5\% at 516~MeV and 10\% above 10~GeV, linearly varying with the logarithm of energy between those values~\citep{Ackermann12}.
We estimated the systematic errors induced in the source flux by repeating the analysis with two sets of modified IRFs where the effective area was scaled up and down by its uncertainty.

In order to evaluate the systematic uncertainties due to the interstellar emission model, we compared the results obtained using the standard model in Section~\ref{image_ana} with the results based on eight alternative interstellar emission models as performed in~\cite{W44_Science} and~\cite{De13}. 
We varied some of the most important parameters of the interstellar emission models, namely the uniform spin temperature used to estimate the column densities of interstellar atomic hydrogen (150~K and 10$^5$~K), the vertical height of the CR propagation halo (4~kpc and 10~kpc), the CR source distribution in the Galaxy (the pulsar distribution by \citealt{Lorimer06} and the SNR distribution by \citealt{Case98}).
We replaced the standard isotropic background and Galactic interstellar emission models with the alternative ones for the spectral analysis. In this procedure, we fixed the isotropic background spectrum while the normalization of the interstellar model components were left free.
The combined systematic errors on the spectral indexes and the spectral shapes considering the above uncertainties are shown in Table~\ref{spec_result} and Figure~\ref{spec}, respectively.

\section{Discussion}
\label{discussion}

\subsection{\A, B, and C}
\label{B_and_C}

Three GeV $\gamma$-ray sources are found to be spatially coincident with the TeV sources \A, 240B, and 240C.
They are likely to be steady sources and overlap with molecular clouds. This suggests that the GeV--TeV $\gamma$-ray emission may be produced by $\pi^0$ decay originating from the interaction between the clouds and CRs which were accelerated in and escaped from the SNR. Alternatively, the three H.E.S.S. sources may be unrelated to W28, but still close to the molecular clouds. Since there is no clear TeV counterpart to \W, we will discuss its origin in the following subsection.

\subsubsection{Individual Known Sources}
First we consider whether the three H.E.S.S. sources may be explained by individual sources unrelated to the runaway CRs from W28.  \A\ is coincident with two \hii~regions, G6.1$-$0.6~\citep{Kuchar97} and G6.225$-$0.569~\citep{Lockman89}, that contain young massive stars which might contribute to the $\gamma$-ray emission.
240B is associated with the clouds containing the ultra-compact \hii~region W28A2, containing a massive star in a very young phase of evolution.
W28A2 exhibits energetic bipolar molecular outflows~\citep{Harvey88,Acord97,Sollins04} which arise from the accretion of matter by the progenitor star. 
\citet{W28}~concluded that GeV $\gamma$-rays can be explained by $\pi^0$-decay and bremsstrahlung. 
The kinetic energy of the outflow is 3.5~$\times$~10$^{46}$~erg and the dynamical timescale is a few 10$^{3}$~yr, with the matter density as high as 10$^{7}$~cm$^{-3}$ \citep{Klaassen06}, although there is no model to explain multi-TeV particle acceleration in such \hii~regions.
240C is spatially coincident with the radio-faint SNR G5.71$-$0.08~\citep{Brogan06}, which was suggested to be interacting with a molecular cloud due to the detection of an OH maser~\citep{Hewitt09}.
Its distance is estimated to be either 3.1 or 13.7~kpc based on the maser velocity. 
Therefore, it is not conclusive that the origins of the H.E.S.S. sources are stellar objects, or otherwise unrelated to SNR W28.

\subsubsection{Cosmic-Ray Escape Model}
We now explore the possibility that the three H.E.S.S. sources are attributed to the $\pi^0$-decay $\gamma$ rays from molecular clouds illuminated by the escaping particles accelerated in W28. 
Attempts to constrain the diffusion coefficient of the runaway CRs have been made recently using the GeV--TeV spectrum of \B~and the TeV data with upper limits in the GeV band~\citep[e.g.,][]{Gabici10,Li10}.
We can expect that the diffusion coefficient may be more tightly constrained by spectral modeling using the GeV--TeV spectra of the H.E.S.S. sources presented in this work.

The CR escape scenario generally assumes that particles accelerated in the SNR are gradually released into the ambient medium. Here we assume an energy-dependent release of accelerated particles after the time $t_{\rm ST}$ when the SNR enters the Sedov phase~\citep[e.g.][]{Gabici09,Ohira11}.
To estimate the $t_{\rm ST}$ of W28, we adopt the following parameters; the typical kinetic energy released by the supernova explosion $E_{\rm SN}~=~{10}^{51}$~erg and the ejecta mass $M_{\rm ej}~=~1.4~M_\odot$.
Assuming evolution in the uniform intercloud gas with a hydrogen number density of 2~cm$^{-3}$~\citep{Reach05}, the Sedov phase started around $t~=~t_{\rm ST} \simeq$~310~yr when the radius of W28 was $r_{\rm ST} \simeq$~2.4~pc.

Let us consider the diffusion process of CRs from SNRs.
We assume that CRs with a momentum $p$ can escape from an SNR at a time $t~=~t_{\rm esc}(p)$ when the SNR radius reaches $R_{\rm esc}(p)$. 
$t_{\rm esc}(p)$ becomes larger for the CRs with lower momentum and is assumed to depend on momentum as a power law, starting with $t_{\rm esc}(p_{\rm max})~=~t_{\rm ST}$ as follows~\citep{Gabici09,Ohira11}:
\begin{eqnarray}
\label{t_esc}
t_{\rm esc}(p)~=~t_{\rm ST}\left(\frac{p}{p_{\rm max}}\right)^{-1/\chi}.
\end{eqnarray}
Using the Sedov-Taylor solution and Equation~(\ref{t_esc}), one finds $R_{\rm esc}(p)~=~r_{\rm ST}(p/p_{\rm max})^{-2/5\chi}$.
Here we adopt the maximum momentum of the particles $p_{\rm max}~=~10^{15}$~eV~c$^{-1}$  (reached at $t_{\rm ST}$) and $\chi$~=~3, following~\citet{Gabici09} and \citet{Ohira11}.
Assuming the age of W28 is 4~$\times$~10$^4$~yr, the SNR is currently releasing CRs with $p~\simeq~$0.5~GeV~c$^{-1}$.

The momentum spectrum of the runaway CRs integrated over the SNR expansion is expected to have the form $N_{\rm esc}(p)~\propto~p^{-s}$ and $s~\sim$~2 if the maximum momentum of the CRs confined in the SNR is a power-law function of time such as Equation~(\ref{t_esc})~\citep[e.g.,][]{Ptuskin05}.
However, the index of the spectrum could be different from $s~=~2$ since it depends on the time history of acceleration efficiency and maximum energy~\citep{Ohira10,Caprioli10}. 
Assuming that particles are injected into DSA from the thermal plasma at the downstream of the SNR shock~\citep{Malkov95} and $\chi~=~3$,  the value of $s$ is changed to be $\sim~2.2$~\citep{Ohira10}. We parameterize the total spectrum of CRs injected into the interstellar space as
\begin{eqnarray}
\label{cr_spec}
N_{\rm esc}(p)~=~k_{\rm esc}p^{-2.2}\exp(-p/p_{\rm max}).
\end{eqnarray}

The spatial distribution of the escaped CRs $n(p,r,t)$ at a time $t$ after the supernova explosion and at a distance $r$ from the SNR center is described by the well-known diffusion equation for the point source case, and it can be solved using the method developed by~\citet{Atoyan95}. Considering the escape from the surface of the expanding SNR shell, \citet{Ohira11} provided the analytical solution of $n(p,r,t)$ as:
\begin{eqnarray}
\label{cr_dist}
n(p,r,t)~=~\frac{N_{\rm esc}}{4\pi^{3/2}R_{\rm d}R_{\rm esc}r}\left[e^{-(r-R_{\rm esc})^2/R_d^2}-e^{-(r+R_{\rm esc})^2/R_d^2}\right],
\end{eqnarray} 
where
\begin{eqnarray}
\label{r_diff}
R_{\rm d}(p,t)~\equiv~2\sqrt{D_{\rm ISM}(p)[t-t_{\rm esc}(p)]}.
\end{eqnarray}
$D_{\rm ISM}(p)$ is the diffusion coefficient of the interstellar medium and is often parameterized with a power-law energy dependence as below:
\begin{eqnarray}
\label{diff_coe}
D_{\rm ISM}(p)~=~10^{27} D_{27}\left(\frac{p}{10~{\rm GeV} c^{-1}}\right)^{\delta}~{\rm cm^2~s^{-1}},
\end{eqnarray}
where $D_{27}$ is the normalization constant.

\subsubsection{Spectral Modeling to Constrain the Diffusion Coefficient}
\label{spectral_modeling}

To minimize the manual scanning of the parameters for constraining of the diffusion coefficient  when modeling the other clouds, we first consider the spectrum of the $\gamma$-ray emission towards the molecular cloud around \B\ because its GeV--TeV spectrum is the best determined among the three sources.
We adopt 1.9~kpc for the distance from the Earth to W28.
Only CR protons are taken into account since the leptonic emission is unimportant in the case of an electron to proton ratio of 0.01, as in the local CR abundance.
The $\gamma$-ray spectrum from $\pi^0$ decays produced by the interaction of protons with ambient hydrogen is scaled by a factor of 1.84 to account for helium and heavy nuclei in the target material and the CR composition~\citep{Mori09}.
The mass of the cloud $M_{\rm B}$ responsible for 240B~is found to be $7.0~\times~10^{4}$~$M_{\odot}$ by the NANTEN CO (J=1--0) data for the velocity range from 0 to 20~km~s$^{-1}$~\citep{Aharonian08}.
Because the distance from the SNR center, $r$, cannot be well determined, we treated it as a free parameter. We adopted the center of the radio boundary shown in Figure~\ref{lat_cmap}(b) as the SNR center.
The minimum $r$ is the projected distance 20~pc derived from the angular distance between the SNR center and the peak of the TeV emission.
From the above assumptions, our model has four adjustable parameters: $r$, $D_{27}$, $\delta$, and $k_{\rm esc}$ which is the normalization of the CR spectrum.

The runaway CR spectrum has two cutoffs whose energies are determined as ($r-R_{\rm esc}$)$^2$/$R_{\rm d}^2$ and ($r+R_{\rm esc}$)$^2$/$R_{\rm d}^2$ from Equation~(\ref{cr_dist}) and $R_{\rm d}$ depends on $D_{27}^{1/2}$ from Equations~(\ref{r_diff}) and (\ref{diff_coe}).
Then, as the value of $D_{27}$ becomes small, the cutoff energies shift to higher energies.
Consequently, for a small value of $D_{27}$, the model spectrum of the $\gamma$-ray emission conflicts with the observed GeV spectrum, and the lower limit obtained is $\sim$~0.5.
In this case, $r$ must be almost the same value as the projected distance, the minimum $r$, because the cutoff energies also move to higher energies with increases in $r$.
On the other hand, $\delta$ is constrained to be 0.20--0.35 to fit the spectrum above the higher cutoff energy as shown in Table~\ref{tab:diff_coe}.
Figure~\ref{spec}(a) shows the $\gamma$-ray model curve with $D_{27}$~=~0.5, $\delta$~=~0.35 and $r$~=~25~pc.
The $\gamma$-ray emission from the Galactic CRs, calculated based on the proton spectrum in \citet{Dermer86}, is also plotted to show the expected background emission. 
Even if the interstellar emission model does not reproduce the background emission from local clouds completely, it is not a large fraction of the observed residual emission.
The upper limit on $D_{27}$ is constrained by the amount of escaped CRs, $W_{\rm p}$, because the normalization of the spectrum of the particles depends on $k_{\rm esc}$/$R_{\rm d} \propto k_{\rm esc}/D_{27}^{1/2}$ from Equations~(\ref{cr_spec}), (\ref{cr_dist}), (\ref{r_diff}) and (\ref{diff_coe}).
Here we assume the upper limit of $W_{\rm p}$ is $10^{50}$~erg, which is 10\% of $E_{\rm SN}$.
We try to fit the $\gamma$-ray spectrum with $D_{27}$~=~1, 5, and 10, respectively. $D_{27}$ is found to be below $\sim$~5 (see Figure~\ref{spec}(a)) since the required $W_{\rm p}$ exceeds the upper limit using $D_{27}$~=~10.
The value of $\delta$ is constrained to be slightly smaller (0.1--0.25) than with $D_{27}$~=~0.5. A similar result was obtained for IC 443 by~\citet{Torres10}.
The variation of $\delta$ for $0.5<D_{27}<5$ is given in Table~\ref{tab:diff_coe}.
The lower limit of $W_{\rm p}$ is obtained with $D_{\rm 27}$~=~0.5 and the minimum $r$, as the spectral normalization is proportional to $k_{\rm esc}/D_{27}^{1/2}r$ from Equations (\ref{cr_spec})--(\ref{diff_coe}).
Using $\delta$~=~0.20, the resulting value of $W_{\rm p}$ is $0.4~\times~10^{49} (M_{\rm B}/7.0~\times~10^{4}~M_{\odot})^{-1}$~erg.

To investigate whether the emission from 240A and 240C can be interpreted within the same CR escape scenario, we modeled the $\gamma$-ray spectra considering the range in the diffusion coefficient obtained for 240B.
The minimum $r$ for 240A and 240C are derived from the projected distances of 21~pc and 24~pc, respectively.
Their spectra can be fitted with models using the same diffusion coefficient, except for the case of $D_{27}$~=~5 (see Table~\ref{tab:diff_coe}).
The model curves with $D_{27}$~=~0.5 and $\delta$~=~0.35 for 240A~and 240C are shown in Figures~\ref{spec}(b) and (c), respectively, with the allowed ranges of $r$ for each combination of the diffusion coefficient and are shown in Table~\ref{tab:diff_coe}.
The masses of the clouds $M_{\rm A}$ and $M_{\rm C}$ towards 240A~and 240C~are estimated to be about 2.3~$\times$~10$^4$~$M_\odot$ and 1.4~$\times$~10$^4$~$M_\odot$ using NANTEN CO (J=1--0) data by \citet{Aharonian08} and \citet{Nicholas12}, respectively.
Using $D_{27}$~=~0.5 and $\delta$~=~0.20, we obtained the lower limits of $W_{\rm p}$ for 240A and 240C as 1.0~$\times$~10$^{49}$($M_{\rm A}$/2.3$~\times$~10$^{4}$~$M_\odot$)$^{-1}$~erg and 1.9~$\times$~10$^{49}$($M_{\rm C}$/1.4~$\times$~10$^{4}$~$M_\odot$)$^{-1}$~erg, respectively.
The different $W_{\rm p}$ resulting from assuming the same diffusion coefficient among the three H.E.S.S. sources may be explained by different fractions of the cloud mass responsible for the $\gamma$-ray emissions.
With $D_{27}$~=~5, $\delta$ is constrained to be 0.1--0.15.  The model curve with $\delta$~=~0.1 is also shown in Figures~\ref{spec}(b) and (c).

By combining the results for the three H.E.S.S. sources, the diffusion coefficient for the runaway CRs from W28 is constrained to be $0.5 \la D_{27} \la 5$ and $0.1 \la \delta \la 0.35$, with a negative correlation between them as shown in Table~\ref{tab:diff_coe}. 
The lower limit of $W_{\rm p}$ is obtained to be 1.9~$\times$~10$^{49}$~erg.
The value of $r$ becomes larger as either $D_{27}$ or $\delta$ increase.
However, all clouds cannot be located beyond three times their projected distances (see Table~\ref{tab:diff_coe}).

\subsubsection{Uncertainty of the Diffusion Coefficient}

We also evaluate the uncertainty of the diffusion coefficient by changing the values of the following parameters; the age of W28, the explosion energy, and the time when the Sedov phase starts.
Doubling the SNR age changes the range of $D_{27}$ to be 0.5--10 while halving changes it to be 0.1--1.
To take into account the uncertainty in $E_{\rm SN}$, we adopt 4~$\times$~10$^{50}$~erg as the lower limit because this value was obtained by \citet{Rho02} with the caution of possible underestimation.
Assuming that the upper limit of $W_{\rm p}$ is 10\% of the $E_{\rm SN}$, $D_{27}$ is constrained to be less than 1. In this case, $\delta$ slightly changes to the harder value of 0.1--0.2 at $D_{27}$~=~1 than those obtained above (see Table~\ref{tab:diff_coe}).
The ambient matter density may also be different from 2~cm$^{-3}$. If W28 originates from a core-collapse supernova, the progenitor makes a cavity into which the SNR initially expands, and the time $t_{\rm ST}$ when CRs start to escape will change.
Adopting a matter density of 0.1~cm$^{-3}$ changes $t_{\rm ST}$ to $\sim$~840~yr . 
Although the spectral break shifts to higher energy, the model curve with each value of $D_{27}$ still does not conflict with the observed spectrum. 
In summary, the diffusion coefficient does not depend strongly on the choice of the parameters.

Moreover, there is uncertainty of the ejecta mass. If we adopt $M_{\rm ej}$~=~7~$M_\odot$ which is likely in core-collapse supernovae, $t_{\rm ST}$ becomes $\sim$~1100~yr which is longer than that with $M_{\rm ej}~=~1.4~M_\odot$.
Consequently, the $\gamma$ rays around 2~GeV could be no longer explained by the CR escape scenario because, based on Equation~(\ref{t_esc}), particles with the momentum lower than about 10~GeV cannot escape.
Thus, it is possible to reproduce all components of the $\gamma$-ray emission if the ejecta mass is small but it would be difficult with masses larger than 7~$M_\odot$.

\subsection{W28}
The $\gamma$-ray emission from the northeastern part of W28 may also originate from the CR escape scenario.  We tried to model the $\gamma$-ray spectrum using the same scenario that we applied to the H.E.S.S. sources. Here, $r$ was fixed to the SNR radius of 13~pc from the radio boundary. The mass of the cloud responsible for the emission $M_{\rm NE}$ is estimated to be 5~$\times$10$^{4}~M_\odot$ by~\cite{Aharonian08}. The model curves with D$_{27}$~=~0.5 and 1 reasonably reproduce the observed spectrum when $\delta$~=~0.35. Thus, the emission is consistent with the CR escape scenario.
However, there are also two other possible scenarios to explain the $\gamma$ rays: (1) CRs remaining at the SNR shell contribute to the emission because this place is the interaction site of W28 with molecular clouds and (2) the emission comes from the reaccelerated CRs in the crushed cloud~\citep{Uchiyama10}.
In the first scenario, $\gamma$ rays above 2~GeV are emitted by particles above 10~GeV~$c^{-1}$, which should have escaped from the SNR shell because $R_{\rm esc}$ for these particles  is $\sim$~11~pc, which is smaller than the SNR radius. However, $R_{\rm esc}$ is proportional to $M_{\rm ej}^{1/3}$ and $n_0^{-1/3}$, and given the uncertainties in these quantities may be larger than 13~pc. Therefore we cannot rule out that these CRs remain within the SNR, and do not use the model results for the northeastern part of W28 to further constrain the diffusion coefficient.

\subsection{\W}
\label{W_and_A}

\W~is separated from the TeV emission, having no obvious counterpart except for the radio emission of W28. 
One possibility is that the $\gamma$ rays come from the particles accelerated and confined in the SNR shell.
The different $\gamma$-ray spectral shape compared with the northeastern boundary of W28, shown in Section~\ref{energy_spectrum}, might be due to the difference of the particle spectrum if the $\gamma$ rays from the northeastern region also come from the CRs confined in the SNR shell.
Since the environment around \W~is expected to be more tenuous than that of the northeastern boundary~\citep[see Figure~2 in][]{Aharonian08}, the effects of damping of magnetohydrodynamic turbulence due to ion-neutral collisions~\citep{Drury96} could operate differently causing a break in the spectrum of accelerated particles at higher energy ($>~100$~GeV). 
A harder radio spectral index is found around \W~\citep{Dubner00}, which may support this scenario.

Another possibility is that the emission originates from the interaction between molecular clouds and the runaway CRs from W28. 
\W~overlaps clouds in the velocity range from 0 to 20~km~s$^{-1}$~\citep[see Figure~2 in][]{Aharonian08}.
The total cloud mass, $M_{\rm W}$, is estimated to be about  3500~$M_\odot$ from the NANTEN CO (J~=~1$-$0) data~\citep{NANTEN1,NANTEN2} assuming the same velocity range and a cloud radius of $\sim$~3~pc derived from the upper limit of the extension of the $\gamma$-ray emission.
We perform modeling of the $\gamma$-ray spectrum in the same manner as in Section~\ref{spectral_modeling}, considering the range of the diffusion coefficient obtained there. 
The lower limit of $r$ is the projected distance of 16~pc.
Using $D_{27}$~=~5, the model cannot reproduce the spectrum as shown in Figure~\ref{spec}(d) and Table~\ref{tab:diff_coe}, because the required $W_{\rm p}$ exceeds 10$^{50}$~erg.
Therefore, $D_{27}$ is constrained to be the smaller range 0.5--1 than that obtained for the three H.E.S.S. sources.
On the other hand, $\delta$ is also tightly constrained to be 0.1--0.25 with $D_{27}$~=~1.
The lower limit of $W_{\rm p}$ is 3.2~$\times$~10$^{49}$($M_{\rm W}$/3.5~$\times$~10$^3$~$M_\odot$)$^{-1}$~erg with $D_{27}$~=~0.5 and $\delta$~=~0.2.
From these results, the four $\gamma$-ray sources around W28 can all be explained by CR escape from the SNR with a single diffusion coefficient with $D_{27}~\simeq~0.5$ and $\delta~\simeq~ 0.35$.

\subsection{Interpretation of the Obtained Diffusion Coefficient}

The obtained values for $D_{27}$ and $\delta$ are smaller than those based on the Galactic CR propagation model: $D_{27}\sim$~10 and $\delta \sim$~0.6~\citep{Ptuskin06,Delahaye08}. 
$D_{27}$ is also smaller than that of W44 derived by~\citet{Uchiyama12}, where is $1 < D_{27} < 30$.
For an SNR in a dense environment, $D_{27}$ is expected to be as low as $\sim$~0.1~\citep{Ormes88} and to depend on the magnetic field strength in the molecular cloud into which the runaway CRs propagate~\citep{Gabici07}.
$D_{27}$ and $\delta$ would also be affected by the amplification of Alfv$\acute{\rm e}$n waves generated by the escaping CRs~\citep{Fujita11}.
The value of $\delta$ depends on the assumption of the spectral index of the accelerated particles in the SNR, which is related to the time history of acceleration efficiency and maximum energy as mentioned above. 
The hard spectrum of the runaway CRs might indicate a harder spectral index than 2.0 for the CRs accelerated in W28 and it is suggested to be 1.7 based on the radio synchrotron spectrum~\citep{Dubner00}.
Non-linear shock modification caused by the efficient CR acceleration can produce such a hard spectrum~\citep[e.g.,][]{Baring99}.

The value of $\delta$ is also strongly dependent on assumptions about the evolution of the accelerator. If an SNR continuously accelerates particles even after the free expansion phase, the spectral index of the escaped particles above the cutoff energy $s'$ is $s$+$\delta$.
This is smaller by a factor of $\delta$/2 than in the case of an impulsive source~\citep{Aharonian08} which is very similar to our assumption for W28, i.e., the value of $\delta$ becomes larger by a factor of 1.5 than our result.
However, it is difficult to adapt this scenario to W28. The SNR is considered to be in the radiative phase with a shock velocity of less than 100~km~s$^{-1}$~\citep{Rho02}, and cannot produce multi-TeV particles.
Therefore, the value of $\delta$ around W28 must be smaller than that of the Galactic CR propagation model.

\section{Conclusion}
\label{conclusion}

We analyzed the GeV $\gamma$-ray emission in the vicinity of SNR W28 using 4 years of LAT data. 
We detected GeV $\gamma$ rays spatially coincident with the TeV sources \A~, B, and C, located south of the radio boundary of W28.
Their spectra in the 2--100~GeV band are consistent with the extrapolation of power-law emission from the TeV $\gamma$-ray sources. 
We also detected GeV emission from \W, located outside the boundary of the TeV emission and coinciding with radio emission from the western shell of W28.
All of the GeV $\gamma$-ray sources overlap with the molecular clouds in the velocity range from 0 to 20~km~s$^{-1}$.

Assuming that the $\gamma$-ray emissions from the three H.E.S.S. sources are due to the decay of $\pi^0$s produced by the interaction of the molecular clouds with CRs escaping from W28, the GeV--TeV spectra can naturally be explained by a single model. We constrain the diffusion constant at 10~GeV~c$^{-1}$ and the power-law index of the energy dependence to be 0.5--5~$\times$~10$^{27}$~cm$^2$~s$^{-1}$ and 0.1--0.35, respectively, with a negative correlation between them. These values are smaller and harder than those of the Galactic CR propagation model.
Considering the masses of the molecular clouds responsible for the emission, the lower limit on the total energy of the escaped CRs is constrained to be $\sim$~2~$\times$~10$^{49}$~erg, in agreement with the conjecture that SNRs are the main sources of the Galactic CRs.
The $\gamma$ rays from \W~can be also interpreted to be the emission originating from the interaction of the runaway CRs and molecular clouds with the same diffusion coefficient as obtained for the H.E.S.S. sources.

\acknowledgments
The research of D.~F.~T. and G.~P. has been done in the framework of the grant AYA2012--39303 and iLINK2011--0303. D.~F.~T. was additionally supported by a Friedrich Wilhelm Bessel Award of the Alexander von Humboldt Foundation. 

The \textit{Fermi} LAT Collaboration acknowledges generous ongoing support
from a number of agencies and institutes that have supported both the
development and the operation of the LAT as well as scientific data analysis.
These include the National Aeronautics and Space Administration and the
Department of Energy in the United States, the Commissariat \`a l'Energie Atomique
and the Centre National de la Recherche Scientifique / Institut National de Physique
Nucl\'eaire et de Physique des Particules in France, the Agenzia Spaziale Italiana
and the Istituto Nazionale di Fisica Nucleare in Italy, the Ministry of Education,
Culture, Sports, Science and Technology (MEXT), High Energy Accelerator Research
Organization (KEK) and Japan Aerospace Exploration Agency (JAXA) in Japan, and
the K.~A.~Wallenberg Foundation, the Swedish Research Council and the
Swedish National Space Board in Sweden.

Additional support for science analysis during the operations phase is gratefully
acknowledged from the Istituto Nazionale di Astrofisica in Italy and the Centre National d'\'Etudes Spatiales in France.

\clearpage

\clearpage

\clearpage

\begin{table}
\begin{center}
\caption{Results of the morphological analysis of the $\gamma$-ray emission from the sources in the vicinity of W28 (2--100~GeV).}
\label{morphology_result}
\begin{tabular}{ccc}
\tableline\tableline
Model & -2$\ln(L_{0}/L)$\tablenotemark{a}  & Additional Degree of Freedom \\ 
\tableline
W28 & 0 & 0\\
Two point sources\tablenotemark{b} & 293.1 & 8  \\
Three point sources\tablenotemark{c} & 334.6 &  12 \\
Four point sources\tablenotemark{d} & 363.5 & 16 \\
H.E.S.S.\tablenotemark{e} + \W & 382.9 & 11\tablenotemark{f} \\ 
\tableline
\end{tabular}
\tablenotetext{a}{2$\ln(L_{0}/L)$, where $L_0$ and $L$ are the maximum likelihood for W28 with the 2FGL disk template and the additional source component, respectively.}
\tablenotetext{b}{\FGLB~and \FGLC~in addition to W28 with positions free in the optimization.}
\tablenotetext{c}{Two point source model plus Source W with optimized position.}
\tablenotetext{d}{Three point source model plus a point source model at the peak of \A.} 
\tablenotetext{e}{H.E.S.S. significance map is used as a template for the intensity of the $\gamma$-ray emission.} 
\tablenotetext{f}{One degree of freedom for the choice of the threshold used for extraction of the regions.}
\end{center}
\end{table}

\begin{table}
\begin{center}
\caption{Power-law spectral indexes and test statistics for the LAT sources near W28 in the 2--100~GeV band.}
\label{spec_result}
\begin{tabular}{ccc}
\tableline\tableline
Name & Index\tablenotemark{a} & TS \\ 
\tableline
\A & 2.12 $\pm$ 0.23 $\pm$ 0.14 & 37 \\
\B & 2.45 $\pm$ 0.19 $\pm$ 0.07 & 88 \\
\C & 2.38 $\pm$ 0.23 $\pm$ 0.17 & 51 \\
\W & 2.06 $\pm$ 0.20 $\pm$ 0.14 & 41 \\
\tableline
\end{tabular}
\tablenotetext{a}{The first and second uncertainties listed represent the statistical and systematic errors, respectively.}
\end{center}
\end{table}

\begin{table}
\begin{center}
\caption{The values of the diffusion coefficient of the escaping CRs and the distance from the SNR center to each molecular cloud obtained by the spectral modeling for \A, B, C, and \W. The spectra of all sources can be reproduced with a single diffusion coefficient with $D_{27}~\simeq~0.5$ and $\delta~\simeq~ 0.35$.}
\label{tab:diff_coe}
\begin{tabular}{cc|cc|cc|cc|cc}
\tableline\tableline
& & \multicolumn{2}{c}{240A} & \multicolumn{2}{c}{240B} & \multicolumn{2}{c}{240C} & \multicolumn{2}{c}{\W} \\ 
\tableline
$D_{27} $ & $\delta$ & $r$ (pc) & $W_{p}$~(10$^{49}$~erg) & $r$ (pc) & $W_{p}$~(10$^{49}$~erg) & $r$ (pc) & $W_{p}$~(10$^{49}$~erg) & $r$ (pc) & $W_{p}$~(10$^{49}$~erg) \\
\tableline
0.5 & 0.35 & 30--40 & 3.3--10 & 25 & 1.0 & 30 & 4.3 & 25 & 10 \\
     & 0.3 & 25--35 & 2.3--4.9 & 20--25 &  0.5--1.0 & 24--30 & 2.1--4.8  & 20--25 & 5.4--10 \\
     & 0.25 & 21--35 & 1.3--4.2 & 20--25 & 0.5--0.8 & 24 & 2.1 & 16--25 &  3.2--7.5 \\
     & 0.2  &  21--30 & 1.0--2.3 & 20  & 0.4 & 24 & 1.9 & 16--25 &  3.2--5.4 \\
     & 0.15 & 21--30 & 0.8--2.0 & N/A  & N/A & N/A & N/A &16--20 & 2.6--3.2 \\
     & 0.1  & 21--25 & 0.7--1.0 & N/A & N/A & N/A & N/A & N/A & N/A \\
\tableline
1 & 0.35 & 40--45 & 8.2--10 & 30--35 & 2.1--2.7 & 35 & 8.1 & N/A & N/A \\
   & 0.3 & 35--45 & 5.9--10 &20--35 & 1.1--2.5 & 24--40 & 4.3--10 & N/A & N/A \\
   & 0.25 & 30--45 & 3.9--8.2 & 20--30 & 0.9--1.6 & 24--35 & 3.2--6.4 & 25 & 10 \\
   & 0.2 & 25--45 & 2.3--7.5 & 20--30 & 0.8--1.5 & 24--35 & 2.7--6.7 &20--25 & 7.5--10 \\
   & 0.15 & 21--40 & 1.6--4.7 &20--25 & 0.6--0.9  & 24--30 & 2.1-3.8 & 16--25 & 5.4--7.5 \\
   & 0.1 & 21--40 & 1.1-4.9 & 20 & 0.5 & 24 & 2.0 &16 & 3.9 \\
\tableline
5 & 0.35 & N/A & N/A & N/A & N/A & N/A & N/A & N/A & N/A \\
  & 0.3 & N/A & N/A & N/A & N/A & N/A & N/A & N/A & N/A \\
  & 0.25 & N/A & N/A & 30--60 & 7.5--10 & N/A & N/A & N/A & N/A \\
  & 0.2 & N/A & N/A & 20--60 & 5.4--10 & 24 & 10  & N/A & N/A \\
  & 0.15 & 30--45 & 6.4 & 20--60 & 4.0--10 & 24--35 & 10 & N/A & N/A \\
  & 0.1 & 21--50 & 4.3--6.4 & 20--45 & 3.2--5.4 & 24--35 & 8.1 & N/A & N/A \\
\tableline
\end{tabular}
\end{center}
\end{table}


\end{document}